\documentclass[%
 reprint,
 amsmath,amssymb,
 aps,
 onecolumn,
]{revtex4-1}

\usepackage{graphicx}
\usepackage{dcolumn}
\usepackage{subfigure}
\usepackage{bm}
\usepackage[utf8]{inputenc}
\usepackage{mathrsfs}
\usepackage[T1]{fontenc}
\usepackage{mathrsfs}
\usepackage{multirow}
\usepackage{amsmath}
\usepackage{amssymb}
\usepackage{graphicx}
\usepackage{subfigure}
\usepackage{esint}
\usepackage{gensymb}
\usepackage{float}
\usepackage{setspace}

\providecommand{\tabularnewline}{\\}
\makeatother
\usepackage{babel}

\begin{document}
\title{High energy beam energy measurement with microwave-electron Compton backscattering}
\begin{spacing}{2.0}
\author{Meiyu Si}
\affiliation{Institute of High Energy Physics, Chinese Academy of Sciences, Beijing 100049, China}
\author{Yongsheng Huang}
\email{ huangysh59@mail.sysu.edu.cn}
\affiliation{School of Science, Campus of Sun Yat-sen University, Shenzhen 518107, China}
\affiliation{Institute of High Energy Physics, Chinese Academy of Sciences, Beijing 100049, China}
\author{Shanhong Chen}
\affiliation{Institute of High Energy Physics, Chinese Academy of Sciences, Beijing 100049, China}
\author{Pengcheng Wang}
\affiliation{Institute of High Energy Physics, Chinese Academy of Sciences, Beijing 100049, China}
\author{Zhe Duan}
\affiliation{Institute of High Energy Physics, Chinese Academy of Sciences, Beijing 100049, China}
\author{Xiaofei Lan}
\affiliation{Physics and Space Science College, China West Normal University, Nanchong 637009, China}
\author{Yuan Chen}
\affiliation{National Synchrotron Radiation Laboratory, University of Science and Technology of China, Hefei 230029, China}
\author{Xinchou Lou}
\affiliation{Institute of High Energy Physics, Chinese Academy of Sciences, Beijing 100049, China}
\affiliation{State Key Laboratory of Particle Detection and Electronics, Institute of High Energy Physics, CAS, Beijing 100049, China}
\author{Manqi Ruan}
\affiliation{Institute of High Energy Physics, Chinese Academy of Sciences, Beijing 100049, China}
\author{Yiwei Wang}
\affiliation{Institute of High Energy Physics, Chinese Academy of Sciences, Beijing 100049, China}
\author{Guangyi Tang}
\affiliation{Institute of High Energy Physics, Chinese Academy of Sciences, Beijing 100049, China}
\author{Ouzheng Xiao}
\affiliation{Institute of High Energy Physics, Chinese Academy of Sciences, Beijing 100049, China}
\author{Jianyong Zhang}
\affiliation{Institute of High Energy Physics, Chinese Academy of Sciences, Beijing 100049, China}
\affiliation{State Key Laboratory of Particle Detection and Electronics, Institute of High Energy Physics, CAS, Beijing 100049, China}

\begin{abstract}
The uncertainty of the energy measurement of the electron beam on circular electron positron collider (CEPC) must be smaller than 10$\mathrm{MeV}$ to make sure the accurate measurement of the mass of the Higgs boson. In order to simplify the energy measurement system, a new method is proposed by fitting the Compton edge of the energy distribution of the gamma ray from a microwave-electron Compton scattering. With our method, the uncertainty of the energy measurement is 6$\mathrm{MeV}$ for the electron energy of $120\mathrm{GeV}$ in the Higgs mode. In this system, the energy resolution of the gamma detection needs to reach $10^{-4}$. Therefore, only the high-purity germanium (HPGe) detector can meet the critical requirement. In a head-on collision mode, the initial photons should be microwave photons with the wavelength of 3.04 centimeters. A cylindrical resonant cavity with selected ${TM_{010}}$ mode is used to transmit microwaves. After the microwave-electron Compton backscattering, the scattered photons and the synchrotron-radiation background transmit a shielding structure and then are detected by a HPGe detector at the end of the beam line of the synchrotron-radiation applications. The hole radius in the side wall of the cavity is about $1.5\mathrm{mm}$ to allow the electron beam passing through. The results of the computer simulation technology (CST) software shows that the influence of the hole radius on the cavity field is negligible. The change of the resonance frequency can be easily corrected by fine-tuning the cavity size. 
\end{abstract}

\maketitle
\section{Introduction\label{sec:Introduction}}
In 2012, the Higgs boson was discovered by ATLAS and CMS at CERN's large hadron collider (LHC)\cite{Aad:2012tfa, Chatrchyan:2012ufa}. A circular electron positron collider (CEPC) was proposed\cite{CEPCStudyGroup:2018ghi} to study the properties of the Higgs boson and to accurately test the predictions of the standard model. The CEPC can also work in W (80$\mathrm{GeV}$) and Z(45.5$\mathrm{GeV}$) modes\cite{CEPCStudyGroup:2018ghi}, looking for clues of the new physics beyond the standard model.

The uncertainty of the measurement of the beam energy on CEPC is required to be on the order of MeV in the Higgs mode. There are two effective methods for the high-precision measurement of the energy of a high-energy electron beam: (1) The resonant depolarization (RDP) method. The resonance depolarization method is used on the large electron positron (LEP) collider for the 45.5$\mathrm{GeV}$ beam energy, and the relative uncertainty is $2\times10^{-5}$\cite{Arnaudon}. However, the RDP method is not proper for the beam energy calibration on CEPC in the Higgs mode because the expected polarization is quite difficult to be obtained in a reasonable time interval. (2) The method of laser-Compton backscattering combining the analysis magnetic field. In this method, the interaction point between the laser pulse and the electron beam is set to be the end of the extraction line of the electron beam and to be front of the deflection magnetic system. After the deflection magnetic field, a long drift vacuum tube is needed. At the end of the optimal drift distance, it needs to be detected that the two-dimensional position distributions of the scattered electron beam and the scattered photons. By analyzing and fitting of the detection data, the uncertainty of the beam-energy measurement can reach $2\mathrm{MeV}$\cite{2020The} in the Higgs mode on CEPC. The precision can meet the physical requirements, but this method has high requirements for the system design and the detectors. It is needed that a separate extraction beam line and several high spatial-resolution and large-area detectors to detect the two-dimensional spatial distributions of the scattered electron beam and the scattered photons.

In the laser-Compton measurement system on Beijing electron positron collider (BEPC II), a $CO_{2}$ laser is employed to collide with the electron beam with the energy of $2\mathrm{GeV}$. The maximum energy of the scattered photons is 6$\mathrm{MeV}$. By fitting the Compton edge of the energy distribution of the scattered photons, the uncertainty of the beam energy measurement can reach $1.2\times10^{-5}$\cite{ZHANG2019391}. If the same method is used in the Higgs mode on CEPC, the energy resolution of the scattered photons needs to reach $10^{-4}$ to meet the measurement accuracy of the beam energy. Only the energy resolution of a high-purity germanium (HPGe) detector can satisfy this requirement. Besides of the restriction of the energy range of a HPGe detector, the accurate online calibration can only be performed for the photon energy smaller than $10\mathrm{MeV}$ by using the $\gamma$-active radionuclides\cite{Muchnoi,gamma-ray}. There are two methods to reduce the photon energy scattered by the electron beam with the energy of 120$\mathrm{GeV}$:(1) Decreasing the collision angle. In order to produce the energy of the scattered photons smaller than $10 \mathrm{MeV}$, the collision angle between the laser pulse and the beam is required to be very small, about $0.04 \mathrm{rad}$. A small collision angle will reduce the interaction luminosity of Compton backscattering and the number of the scattered photons. The repeatability and stability of the small collision angle are difficult to be realized. (2) Replacing the $CO_{2}$ laser to a longer-wavelength light source. In a head-on collision mode, a microwave beam with the wavelength of a few centimeters can be employed to collide with an electron beam with the energy of 120$\mathrm{GeV}$ to obtain the scattered photons with the energy smaller than $10 \mathrm{MeV}$.

Therefore, a new microwave-electron Compton backscattering method is proposed to measure the high energy beam energy. A cylindrical resonant cavity with ${TM_{010}}$ mode is selected to store the energy of the microwave photons. The microwave photons and electrons undergo head-on Compton backscattering, and then pass through a magnet separation system. At the end of the synchrotron-radiation beam line, before the scattered photons and the synchrotron radiation photons are detected by a HPGe detector, a shielding structure is designed to minimize background noises, such as low-energy synchrotron radiations and the classical radiation from the electron beam in the cavity. This method reduces the complexity of the system design and the cost. The maximum energy of the scattered photons is chosen to be higher than the better calculated energy to reduce the influence of the synchrotron radiation background. In consequence to a head-on collision mode, the maximum energy of the scattered photons is determined to be $9 \mathrm {MeV}$. The energy of the initial photons is calculated to be $4.08\times10^{-5} \mathrm{eV}$ by the Compton backscattering process. The corresponding wavelength of the initial photons is 3.04$\mathrm{cm}$. With our method, the measurement uncertainty of the electron beam with the energy of $120\mathrm{GeV}$ in the Higgs mode on CEPC is $6\mathrm{MeV}$. 
In Sec. $\MakeUppercase{\romannumeral 2}$, the principle of the Compton backscattering process, parameters choice of the microwave and the system design are introduced. In Sec. $\MakeUppercase{\romannumeral 3}$, the number of the scattered photons is obtained by calculating Compton backscattering cross section and luminosity. In Sec. $\MakeUppercase{\romannumeral 4}$, a shielding structure with $400\mathrm{cm}$ polyethylene and $0.2\mathrm{cm}$ lead are added to minimize the background noise. The influence of the hole radius on the cavity field and the influence of the hole radius on the resonance frequency are simulated by the CST software. In Sec. $\MakeUppercase{\romannumeral 5}$, discussions on the systematic errors of the beam-energy measurement are given.
 
\section{Interaction process\label{sec:Interaction process}}
\subsection{Compton backscattering.}
In 1948, Feenberg and Primakoff proposed the kinematics formula of the Compton scattering process\cite{1948Interaction}. The conservation of the four momentum can be written as
\begin{equation}
p+k_{0}\to p^{'}+k,
\end{equation}
where p=($\varepsilon_{0}$, $\vec p$), $k_{0}$=($\omega_{0}$, $\vec k_{0}$) is the four momentum of the initial electrons and the initial photons respectively;
$p^{'}$=($\varepsilon $, $\vec p^{\prime}$)
, $k$=($\omega^{\prime}$, $\vec k$) is the four momentum of the scattered electrons and the scattered photons respectively. Figure 1 shows the Compton backscattering process. The electrons and photons collide at an angle of $\alpha$ to produce the scattered electrons and the scattered photons. 
\begin{figure}[ht]
\centering\includegraphics[scale=0.8]{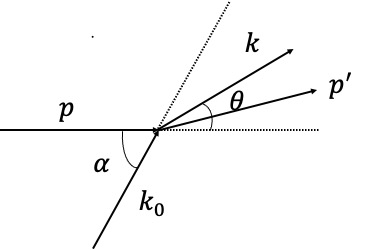}
\caption{The p=($\varepsilon_{0}$, $\vec p$), $k_{0}$=($\omega_{0}$, $\vec k_{0}$) is the four momentum of the initial electrons and the initial photons respectively; $p^{'}$=($\varepsilon $, $\vec p^{'}$), $k$=($\omega^{\prime}$, $\vec k$) is the four momentum of the scattered electrons and the scattered photons respectively. The electrons and photons undergo Compton backscattering at the collision angle $\alpha$ to produce the scattered electrons and the scattered photons. $\alpha$ is the collision angle between the initial electrons and the initial photons momenta. $\theta$ is the angle between the initial electrons and the scattered photon momenta.  \label{fig:Compton backscattering process} }
\end{figure}
For two-body scattering, the Mandelstam variable is defined as
\begin{equation}
\begin{split}
s=(p+k_{0})^{2}=(p^{'}+k)^{2}=m^{2}+2\varepsilon_{0}\omega_{0} (1-\beta \mathrm{cos}\alpha),\\
t=(p^{'}-p)^{2}=(k-k_{0})^{2}=-2\omega_{0}\omega^{\prime}(1-\mathrm{cos}(\alpha-\theta)),\\
u=(p-k)^{2}=(k_{0}-p^{'})^{2}=m^{2}-2\omega^{\prime}\varepsilon_{0}(1-\beta \mathrm{cos} \theta),
\end{split}
\end{equation}
where $\beta$ is the velocity of the electrons, $m$ is the electron rest mass expressed in energy units. Substituting Eq.(2) into $s+t+u=2m^{2}$ and appropriately simplifying, the energy of the scattered photons is given by
\begin{equation}
	\omega^{\prime}=\frac{\omega_{0}(1-\beta \mathrm{cos} \alpha)}{1-\beta \mathrm{cos} \theta +\frac{\omega _{0}}{\varepsilon_{0}}(1-\mathrm{cos}(\alpha -\theta))}.
\end{equation}
For $\theta =0$, the maximum energy of the scattered photon is
\begin{equation}
	\omega_{max}^{\prime}=\frac{\omega_{0}(1-\beta \mathrm{cos} \alpha)}{1-\beta+\frac{\omega _{0}}{\varepsilon_{0}}(1-\mathrm{cos} \alpha)}.
\end{equation}
For $\varepsilon_{0}\gg m\gg \omega _{0}$, $\beta =\sqrt{1-m^{2}/\varepsilon_{0}^{2}}\approx 1-m^{2}/2\varepsilon_{0}^{2}$. Equation (4) can be simplified to be\cite{2008The}
\begin{equation}
\omega_{max}^{\prime}=\frac{\varepsilon_{0}^{2}\mathrm{sin}^{2}(\frac{\alpha}{2})+\frac{m^{2}}{4}\mathrm{cos}\alpha}{\varepsilon_{0}\mathrm{sin}^{2}(\frac{\alpha}{2})+\frac{m^{2}}{4\omega_{0}}}.
\end{equation}
If $\alpha=\pi$, then
\begin{equation}
\omega_{max}^{\prime}=\frac{\varepsilon_{0}^{2}-\frac{m^{2}}{4}}{\varepsilon_{0}+\frac{m^{2}}{4\omega_{0}}}.
\end{equation}
The wavelength of the scattered photons can be obtained by $\omega_{max}^{\prime}=\frac{hc}{\lambda}$, where h=4.13566743$\times 10^{-15} \mathrm{eV/s}$, $c=3\times10^{8}\mathrm{m/s}$. Using Eq.(6), the beam energy can be written as follows:
\begin{equation}
\varepsilon_{0}=\frac{\omega_{max}^{\prime}}{2} \left(1+\sqrt{1+\frac{m^{2}}{\omega_{0} \omega_{max}^{\prime}}}\right).
\end{equation}
The high energy beam-energy measurement with microwave-electron Compton backscattering is applied to the Higgs mode on CEPC. Table 1 shows the parameters of the beam\cite{CEPCStudyGroup:2018ghi}.
\begin{table}[h]
\centering
\begin{tabular}{cc}
\hline
  Parameters in the Higgs mode & Values \tabularnewline
\hline
 Beam energy $\varepsilon_{0}$ (GeV) & 120\tabularnewline
 Bunch number B & 242  \tabularnewline
 Particles/bunch $N_{2} (10^{10})$  & 15  \tabularnewline
 Revolution frequency $f^{\prime} (s^{-1}) $  & 3000  \tabularnewline
 Bunch spacing ($\mathrm{ns}$) & 680 \tabularnewline
 Beam current $I$ ($\mathrm {mA}$) & 17.4 \tabularnewline
 Bending radius $\rho$ (km) & 10.7 \tabularnewline
 Beam size $\sigma_{x_{2}}/\sigma_{y_{2}}(\mathrm{\mu m})$ & 200-450/5-20\tabularnewline
 Bunch length $\sigma_{z_{2}} (\mathrm{mm})$ & 4.4 \tabularnewline
\hline 
\end{tabular}
\caption{The beam parameters in the Higgs mode on CEPC. Taking the beam size $\sigma_{x_{2}}=300\mathrm{\mu m},  \sigma_{y_{2}}=15\mathrm{\mu m}$ for theoretical calculation.}
\end{table}
The energy spectrum of the scattered photons observed experimentally has a cutoff point at the maximum energy, known as the Compton edge. The maximum energy of the scattered photons is obtained by fitting the Compton edge. Therefore, the beam energy is determined by Eq.(7). The number of the scattered photons per second in Compton backscattering satisfies the following relationship
\begin{equation}
\frac{dN_{\gamma}}{dt}=L\sigma.
\end{equation}
$L$ is the luminosity, which determines the ability of the collider to generate events. $\sigma$ is the cross section. After differentiating Eq.(8) with respect to the scattered-photon energy, $\omega^{\prime}$, it is obtained:
 \begin{equation}
\frac{dN_{\gamma}}{d\omega^{\prime} dt}=L\frac{d\sigma}{d\omega^{\prime}},
\end{equation}
where the $\frac{d\sigma}{d\omega^{\prime}}$ is the interaction differential cross section. The number of the scattered photons can be obtained after calculating the differential cross section and the luminosity. The detail calculation process will be discussed in Sec. $\MakeUppercase{\romannumeral 3}$.

\subsection{Resonant cavity.}
With our method, the uncertainty of the energy measurement is 6$\mathrm{MeV}$ for the electron energy of 120$\mathrm{GeV}$ in the Higgs mode. In this system, the energy resolution of the gamma detection needs to reach $10^{-4}$. Only a HPGe detector can reach the required energy resolution and also restrict the maximum energy of scattered-photons to be smaller than 10$\mathrm{MeV}$\cite{gamma-ray}. The maximum energy of the scattered photons is chosen to be higher than the better calculated energy to reduce the influence of the synchrotron radiation background. Consequently, the maximum energy of the scattered photons is selected to be 9$\mathrm{MeV}$. From Eq.(6), for 120$\mathrm{GeV}$ electrons on CEPC, the initial photons should be microwave photons with the wavelength of 3.04 centimeters. In the $TM_{010}$ resonant mode, the wave number $K=K_{c}=\frac{V_{01}}{R}=\frac{2.405}{R}$, where R is the radius of the cavity, $V_{01}$ is the first root of the zero-order Bessel function. The resonance frequency $f=K\cdot c/2\pi$ and the resonance wavelength $\lambda=2\pi/K=2.613R$. The radius $R = 11.65 \mathrm{mm}$ of the resonant cavity can be determined. The length of the cavity is selected to be $20\mathrm{mm}$, because the $TM_{010}$ is the main mode when the cavity length $l<2.1R$. Figure 2 shows the resonant cavity and the relative parameters. There are three coordinate systems, including the cylindrical coordinate system ($r_{1}$,$\varphi_{1}$,$z_{1}$) of the resonant cavity, the rectangular coordinate system ($x_{2},y_{2},z_{2}$) of the moving electron beam, the rectangular coordinate system ($ x,y,z$) used to calculate the interaction process. The electron beam passes through the resonant cavity and undergoes Compton backscattering with the microwave photons.
\begin{figure}[ht]
\centering\includegraphics[scale=0.4]{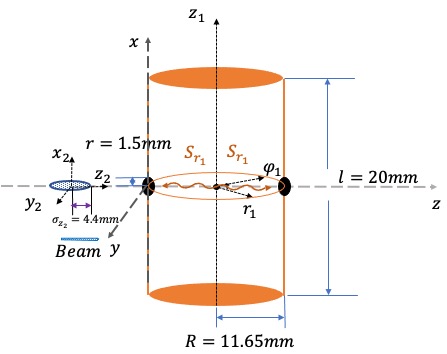}
\caption{The electron beam passes through the resonant cavity and undergoes Compton backscattering with the microwave photons. The $S_{r_{1}}$ is the energy-flux density of the microwave photons. The direction of the microwave photons is in the radial direction of the cavity.}
\end{figure} 

For the $TM_{010}$ mode in the cylindrical coordinate system, the electromagnetic field is rotationally symmetric in the cavity\cite{Microwave2}. The electromagnetic field in the resonant cavity is given by:
\begin{equation}
E_{z_{1}}=E_{m}J_{0}(K_{c}r_{1})e^{j\omega t},
\end{equation}
\begin{equation}
H_{\varphi_{1}}=jE_{m}\frac{1}{\eta}J_{1}(K_{c}r_{1})e^{j\omega t},
\end{equation}
\begin{equation}
E_{r_{1}}=E_{\varphi_{1}}=H_{r_{1}}=H_{z_{1}}=0,
\end{equation}
where $\omega$ is the angular frequency, $\eta=\sqrt {\frac {\mu_{0}}{\varepsilon_{0}}}$, the vacuum permeability $\mu_{0}=4\pi\times10^{-7}\mathrm{H/m}$,  the vacuum dielectric constant $\varepsilon_{0}=8.854188\times10^{-12}\mathrm{F/m}$. Combining Eq.(10), Eq.(11) and Eq.(12), the electric field has only the longitudinal field in the $z_{1}$ direction, the magnetic field has only the transverse field in the $\varphi_{1}$ direction. 
Figure 3 shows the distribution of the electric field in the transverse section and the distribution of the magnetic field in the longitudinal section of the resonant cavity. The electric field is in the $z_{1}$ direction, and decreases with the radius. The magnetic field is in the $\varphi_{1}$ direction, and increases with the radius. The amplitude corresponds to the microwave energy stored in the cavity. 
\begin{figure}[ht]
  \centering
  \subfigure{\includegraphics[width=3.3in]{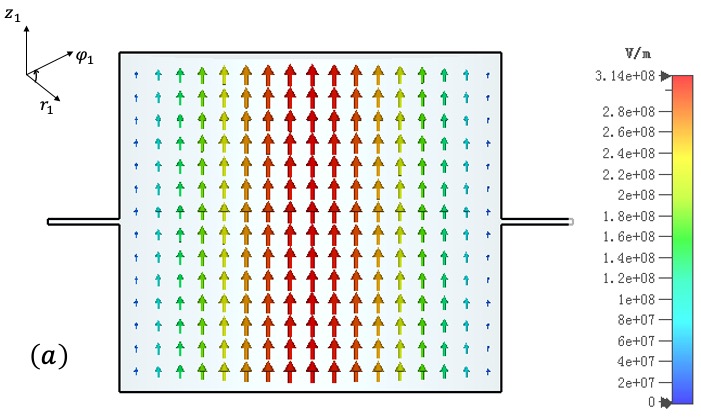}}
  \subfigure{\includegraphics[width=2.8in]{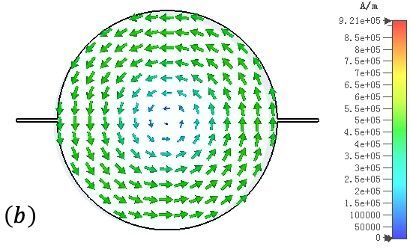}}
  \caption{The distribution of the electric field in the transverse section and the distribution of the magnetic field in the longitudinal section of the resonant cavity. (a) The electric field in the resonant cavity is in the $z_{1}$ direction, and decreases with the radius. (b) The magnetic field is in the $\varphi_{1}$ direction, and increases with the radius.}
\end{figure}
The Poynting vector is the energy-flow density vector in the electromagnetic field. According to the electromagnetic field, the Poynting vector is 
\begin{equation}
\overrightarrow{S}=-Re(E_{z_{1}})\times Re(H_{\varphi_{1}})\overrightarrow{r_{1}} =\frac{1}{\eta}{E_{m}^{2}J_{0}(K_{c}r_{1})J_{1}(K_{c}r_{1})\mathrm{sin}(\omega t)\mathrm{cos}(\omega t)}\overrightarrow{r_{1}},
\end{equation}
where $Re(.)$ stands for the real part. Set $\overrightarrow{S}=S_{r_{1}}\overrightarrow{r_{1}}$, the $S_{r_{1}}$ represents the energy per second passing through a vertical unit area. Figure 4 shows the distribution of the Poynting vector, which is propagating from the central axis $z_{1}$ along the radius to the cavity wall. The field intensity of $E_{m}$ is chosen to be $1\times10^{7} \mathrm{V/m}$. The oscillation period is $50\mathrm{ps}$.
\begin{figure}[ht]
\centering\includegraphics[scale=0.4]{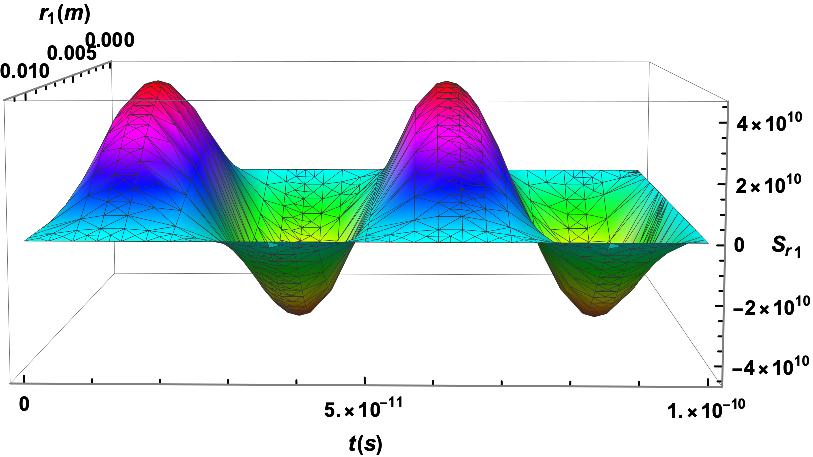}
\caption{The Poynting vector is propagating from the central axis $z_{1}$ along the radius to the cavity wall. The oscillation period $T= 50 \mathrm{ps}$.\label{fig:Three-dimensional variation} }
\end{figure}

The direction of the microwave photons is the same as the $\overrightarrow{S}$. The $S_{r_{1}}<0$ means that the microwave photons are moving in the opposite direction. As shown by Figure 2, the holes are made in the side wall of the cavity to make sure the electron beam passing through. Table 1 shows the bunch size $\sigma_{x_{2}}=300\mathrm{{\mu}m}$, $\sigma_{y_{2}}=15\text{\ensuremath{\mu}m}$, $\sigma_{z_{2}}=4.4\mathrm{mm}$. The radius of the holes is adjustable within a certain range. If the hole radius is too small, the beam will hit the side wall of the cavity. The jitter up and down of the electron beam needs to be considered, and the radius of the hole is generally $3\sigma_{x_{2}}$ to $5\sigma_{x_{2}}$. The hole radius is $5\sigma_{x_{2}}$ about $1.5\mathrm{mm}$ to allow the electron beam passing through. In Sec. $\MakeUppercase{\romannumeral 5}$, it will be discussed that the influence of the hole radius on the cavity field and the influence of the hole radius on the resonance frequency.

\subsection{System design.}
Figure 5 shows the system design of the microwave-electron Compton backscattering. The Compton backscattering occurs when the electron beam passes through the resonant cavity. The design is mainly divided into three parts: (1)the first is the Compton backscattering process of the electron beam and the microwave photons; (2)the second is the magnet separation system of the scattered photons and the scattered electrons. At the same time, the synchrotron radiations are emitted due to the deflection of the electron beam in the magnetic field; (3)the gamma-ray detection system. At the end of the beam line, a shielding structure is designed to minimize background noises, such as low-energy synchrotron radiations and so on. After that, the energy of the scattered photons is detected by the HPGe detector.
\begin{figure}[ht]
\centering\includegraphics[scale=0.5]{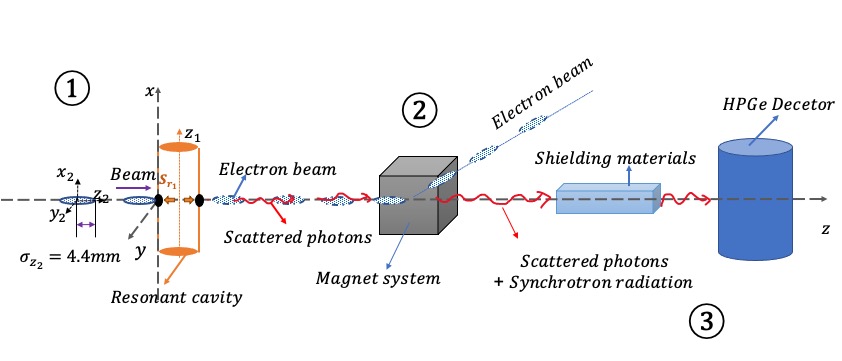}
\caption{The system design of the microwave-electron Compton backscattering. The design is mainly divided into three parts:(1)the first is the Compton backscattering process of the electron beam and the microwave photons; (2)the second is the magnet separation system of the scattered photons and the scattered electrons. At the same time, the synchrotron radiations are emitted due to the deflection of the electron beam in the magnetic field.; (3)the gamma-ray detection system. A shielding structure is designed to minimize background noises, such as low-energy synchrotron radiations and so on. Then the HPGe detector is used to detect the energy of the scattered photons.\label{fig:Simple model}}
\end{figure}
Figure 6 is an electron and photon separation device designed for synchrotron radiation applications on CEPC. The output distance is 700 meters. There is enough space at the end of the synchrotron radiation application to design the shielding. The resonant cavity is placed in front of the separation system. Therefore, our method for the beam-energy measurement does not need to set up a new extraction system of the scattered photons or the scattered electron beam, and reduces the complexity of the system design and the cost. Our method also provides an important cross-check way for the laser-Compton method\cite{2020The}. The design of the shield, including materials and dimensions will be given in the Sec. $\MakeUppercase{\romannumeral 4}$.
\begin{figure}[ht]
  \centering
  \subfigure{\includegraphics[width=2.8in]{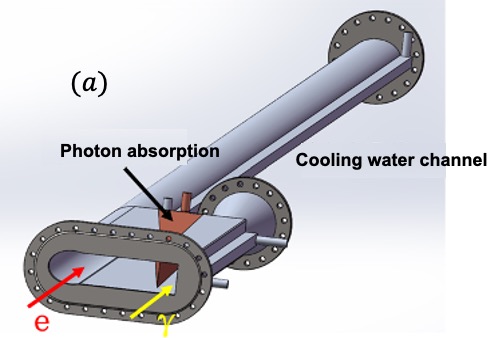}}
  \subfigure{\includegraphics[width=3.6in]{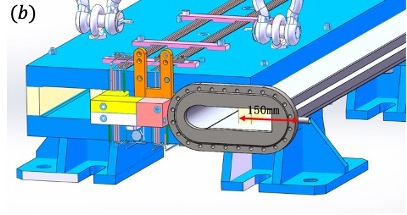}}
  \caption{ The electrons and photons separation device designed for the beam line of the synchrotron radiation applications on CEPC. Figure 6(a) shows the end part of the enlarged vacuum chamber shown in Figure 6(b).}
\end{figure}

\section{Calculation process\label{sec:Interaction process}}
\subsection{Differential cross section.}
In a free space, the photons propagate in the form of a plane electromagnetic wave. Since the cylindrical resonant cavity is bounded, the existence of the boundary conditions creates a standing wave field in the cavity. The Compton backscattering differential cross section in the resonant cavity is a little different from that in a free space. In the reference \cite{cross-section}, the differential cross section in the resonant cavity is equivalent to adding the coefficient $|F_{(TM010)}^{(1)}|^{2}$,
\begin{equation}
\frac{d\sigma_{0}}{d\omega^{\prime}}=|F_{(TM010)}^{(1)}|^{2}\cdot 2\pi\frac{r_{e}^{2}}{\kappa^{2}(1+u)^{3}}\frac{\varepsilon_{0}}{(\varepsilon_{0}-\omega^{\prime})^{2}}\{\kappa[1+(1+u)^{2}]-4\frac{u}{\kappa}(1+u)(\kappa-u)\},
\end{equation}
where $\kappa(\alpha)=4\frac{\omega_{0}\varepsilon_{0}}{(mc^{2})^{2}}\mathrm{sin}^{2}(\frac{\alpha}{2})$, the dimensionless constant $u=\frac{\omega^{\prime}}{\varepsilon_{0}-\omega^{\prime}}$.
The maximum energy of the scattered photons is $9\mathrm{MeV}$ and the interaction angle $\alpha=\pi$, the wavelength of the initial microwave photons is $3.04\mathrm{cm}$. The differential cross section is about $0.04 \mathrm{barn}$\cite{cross-section}. To obtain the number of the scattered photons, it is necessary to calculate the interaction luminosity, which is related to the number of the initial electrons and the initial microwave photons.

\subsection{Luminosity and the number of the scattered photons.}
The Poynting vector $\overrightarrow{S}$ is divided by the energy of a single microwave photon $\omega_{0}$ to get the areal density of the photon number as follows
\begin{equation}
\overrightarrow{\sigma_{m}}=\frac{\overrightarrow{S}}{\omega_{0}}=\frac{1}{\eta \omega_{0}}{E_{m}^{2}J_{0}(K_{c}r_{1})J_{1}(K_{c}r_{1})\mathrm{sin}(\omega t)\mathrm{cos}(\omega t)}\overrightarrow{r_{1}},
\end{equation}
where the unit is $1/\mathrm{(m^{2}\cdot s)}$. 
The interaction luminosity of the electron beam and the microwave photons can be calculated by \cite{T1976}
\begin{equation}
L=N_{2}\cdot 2Bf^{'}\int \sigma_{m}(r_{1})f_{2}(x_{2},y_{2},z_{2},t)dxdydzdt,
\end{equation}
where $N_{2}$ is the number of the electrons in a single bunch, B is the number of bunches, $f^{'}$ is the revolution frequency. For the single effect of a single electron bunch, $B=1,f^{'}=1$.
The normalization density function of one electron bunch, $f_{2}$, can be written as  
\begin{equation}
f_{2}(x_{2},y_{2},z_{2},t)=\frac{1}{2\pi\sigma_{x2}\sigma_{y2}\cdot \sqrt{2\pi}\sigma_{z2}}\mathrm{exp}\left[-\frac{1}{2}\left(\frac{x_{2}^{2}}{\sigma_{x2}^{2}}+\frac{y_{2}^{2}}{\sigma_{y2}^{2}}+\frac{z_{2}^{2}}{\sigma_{z2}^{2}}\right)\right],
\end{equation}
where $\int {f_{2}(x_{2},y_{2},z_{2},t)} {dx_{2}} {dy_{2}} {dz_{2}}=1$. The parameters can be obtained from Table 1.
The dimension of the luminosity is $[1/m^{2}]$, which represents the number of particle collisions per unit area for a single electron bunch.The integration range is determined by the size of the electron bunch.

Figure 7 shows the interaction processes between an electron beam and the microwave photons in the cavity. $t_{0}$ is defined as the moment when the beam reach the left wall of the cavity. The oscillation period of the microwave photons is $50\mathrm{ps}$. In the left side of $z_{1}$, the beam and the microwave photons move in a head-on manner, and the beam will experience two complete wave packets. In the first wave packet, the overlapping length between the beam and the microwave photons is $5.626 mm$($c\cdot t_{1}$); in the second wave packet, the overlapping length is an electron bunch length of $8.8 \mathrm{mm}$ as shown in Table I. In the right side of $z_{1}$, the beam and microwave photons move in the same direction at the speed of light $c$. For $\sigma_{m} < 0 $, the head-on collision occurs. The overlapping length of the beam and the microwave photons is $5.509 mm$ ($c\cdot(t_{4}-t_{3})$). Until it passes through the cavity at $t_{6}$.

\begin{figure}[ht]
\centering\includegraphics[scale=0.5]{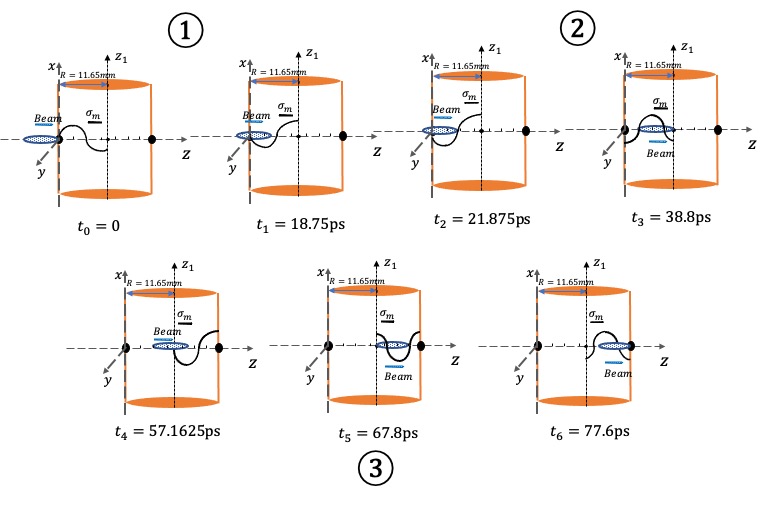}
\caption{The interaction processes between an electron beam and the microwave photons in the cavity. The oscillation period of the microwave photons is $50\mathrm{ps}$. In the left side of $z_{1}$, the beam and the microwave photons move in a head-on manner, and the beam will experience two complete wave packets. In the right side of $z_{1}$, the beam and the microwave photons move in the same direction at the speed of light $c$. For $\sigma_{m} < 0 $, the head-on collision occurs. \label{fig:A single electron bunch passing through the cavity.} }
\end{figure}
For $\omega_{max}^{\prime}=9\mathrm{MeV}$, the luminosity of the three parts is $4.30\times10^{33}/\mathrm{m^{2}}$,  $5.14\times10^{33}/\mathrm{m^{2}}$, $3.18\times10^{33}/\mathrm{m^{2}}$ respectively with Eq.(16). The differential cross section is $0.04 \mathrm{barn}$ with Eq.(14). According to Eq.(9), the number of the scattered photons in the three parts is $17193$, $20541$, $12725$ respectively. Therefore, the total number of the scattered photons can reach 50459 in a total interaction process between an electron beam with the microwave photons in the cavity.
$N_{e}$ is the number of the electrons, $N_{mp}$ is the number of the microwave photons, $\sigma$ is the cross-section of the Compton backscattering. The number of the scattered photons can be estimated by $N_{\gamma}=N_{e}\cdot N_{mp}\cdot \sigma/S$, where S is the cross-sectional area of interaction. As shown in Figure 7, the number of the scattered photons is calculated in three parts. For example, in the first part, $N_{mp}=1.4\times10^{15}$ and $N_{e}=2.8\times10^{10}$ by integral, the $N_{\gamma}=10945$. Therefore, the number of the scattered photons that come out from the three parts is 66896.

\section{Simulation process \label{sec:Simulation process}}
\subsection{Synchrotron radiation.}
As the electron velocity approaches the speed of the light, the angular distribution of the synchrotron radiation shrinks within a very narrow range in the direction of the electron velocity. The radiation angle of a high-speed moving charged particle is estimated to be
\begin{equation}
	\theta=\frac{1}{\gamma}\approx \frac{mc^{2}}{\varepsilon_{0}}.
\end{equation}
For the 120$\mathrm{GeV}$ beam energy, the divergence angle is $0.0043 \mathrm{mrad}$. The synchrotron radiation will travel more than a thousand meters to reach a range of $1\mathrm{cm}$, and the main intensity will still be distributed around an angle of 0. Therefore it is impossible to distinguish spatially the synchrotron photons and the Compton scattered photons on a HPGe detector. The synchrotron radiation spectrum is continuous, evenly in the horizontal direction, with the strongest at the central plane in the vertical direction. At the central plane of the radiation, the intensity is integrated in the vertical direction to obtain the photon flux $F_{bm}$ as follows\cite{SR}: 
\begin{equation}
\frac{dF_{bm}(y)}{d\theta}=2.457\times10^{13}E(\mathrm{GeV})I(A)G(y).
\end{equation}
where the unit of the photon flux is $photons\mathrm{/s/mrad/0.1\%BW}$, $\theta$ is the horizontal observation angle. $I$ is the storage ring current, $I=17.4 \mathrm{mA}$. $G(y)$ is expressed by
\begin{equation}
G(y)=y\int_{y}^{\infty}K_{\frac{5}{3}}(y^{'})dy,
\end{equation}
where K is the second kind of modified Bessel function. $y=\epsilon/\epsilon_{c}$. $\epsilon$ is the photon energy. $\epsilon_{c}$ is the critical photon energy written as follows:
\begin{equation}
\epsilon_{c}=2.218\frac{\varepsilon_{0}^{3}}{\rho},
\end{equation}
where $\varepsilon_{0}$ is the energy of electrons, $120 \mathrm{GeV}$ in the Higgs mode. With Table 1, the bending radius $\rho$ is equal to $10.7 \mathrm{km}$. Therefore, the critical photon energy is $358.2\mathrm{keV}$.
Considering the single effect of a single bunch and the horizontal observation  angle, $0.2\mathrm{m rad}$, dividing by the bunch number $B=242$ and the revolution frequency $f^{\prime}=3000 s^{-1}$, the flux spectrum of the synchrotron radiation can be obtained and shown in Figure 8.
\begin{figure}[ht]
  \centering
  \subfigure{\includegraphics[width=3.25in]{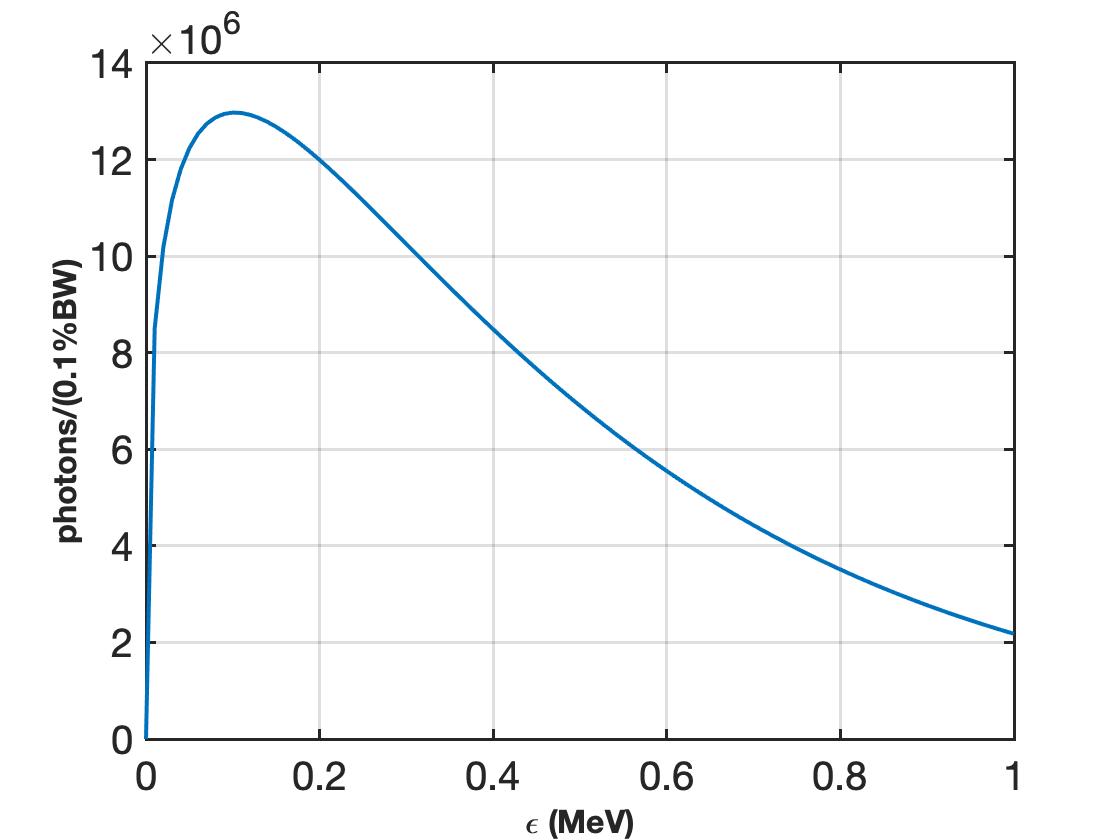}}
  \subfigure{\includegraphics[width=3.25in]{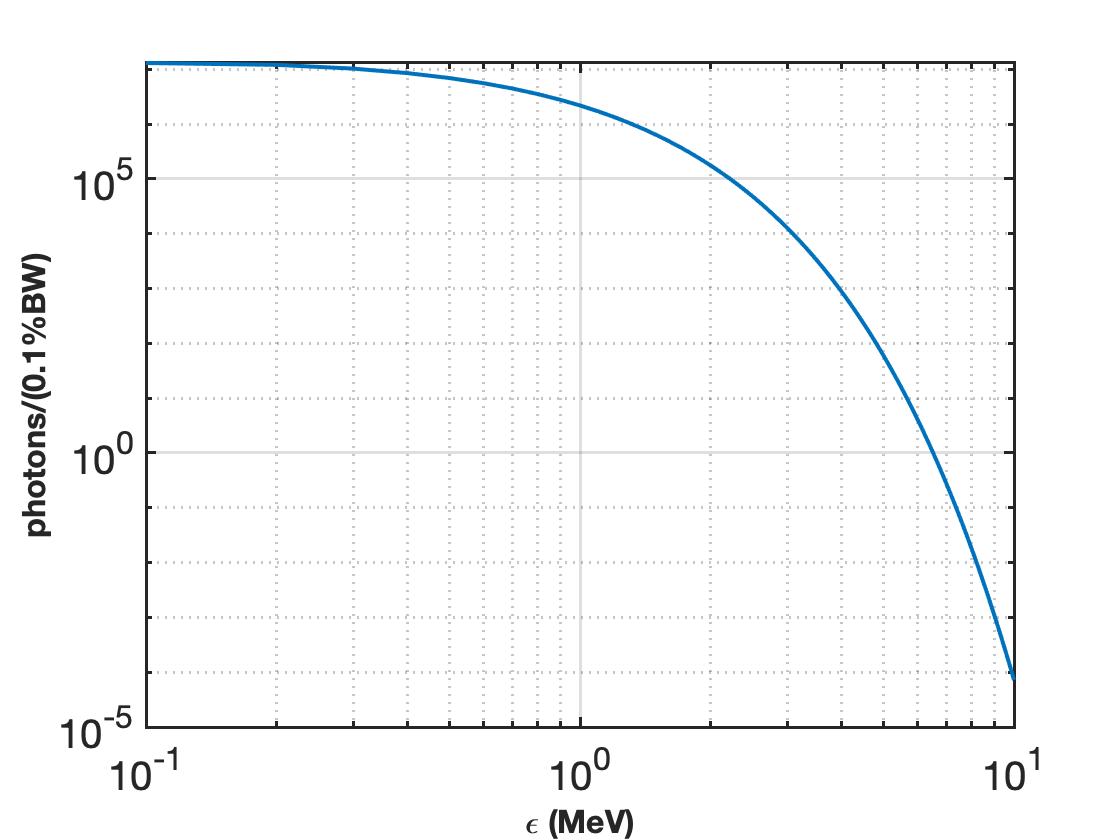}}
  \caption{ The flux spectrum of the synchrotron radiation. $\epsilon$ is the photon energy. The vertical axis is the photon flux, the unit is $photons/\mathrm{0.1\%BW}$. (a) The energy range of the photons is from 0 to $1\mathrm{MeV}$; (b) The energy range of the photons is from 0 to $10\mathrm{MeV}$.}
 \end{figure}
 
The acceptance rejection sampling method\cite{sampling} is used in the differential cross section formula of Eq.(14), so as to extract the value of scattering parameter $u$. Combining the definition of $u$ and the number of scattered photons at 9$\mathrm{MeV}$, the energy $\omega^{\prime}=\frac{u}{1+u}\varepsilon_{0}$ of the scattered photons is obtained. Figure 9 shows the energy spectrum after the acceptance rejection sampling.
\begin{figure}[ht]
\centering\includegraphics[scale=0.4]{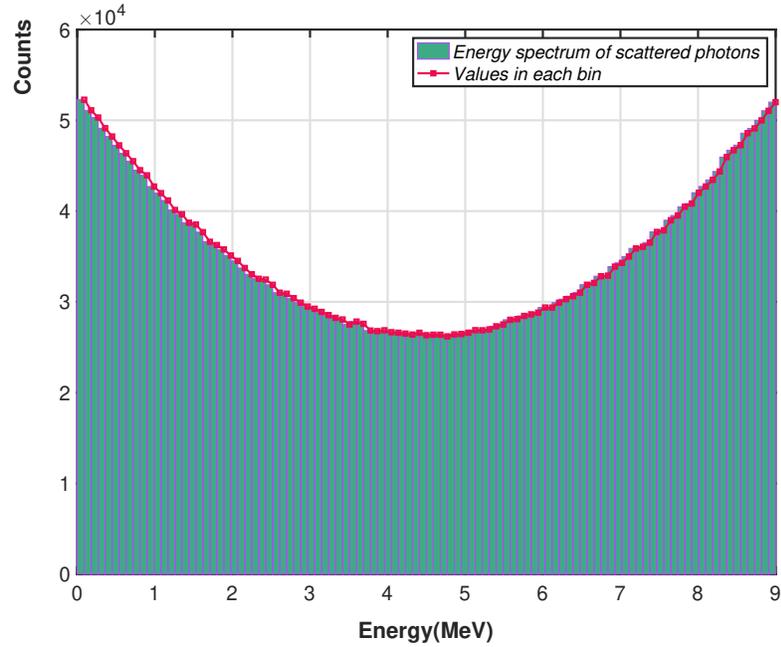}
\caption{ The energy spectrum of the scattered photons after the acceptance rejection sampling. The red line is the values in each bin.\label{fig:The acceptance rejection sampling} }
\end{figure}
A shielding structure is designed with the help of Monte Carlo simulation based on the Geant4 package to minimize background noises, such as low-energy synchrotron radiations photons and so on. Figure 10 shows the input source for Geant4 simulation, including the scattered photons and the synchrotron radiation.  The inset in Figure 10 shows these two parts respectively, the black triangle represents the synchrotron radiation from Figure 8(b), the red rectangle represents the energy spectrum of the scattered photons from Figure 9. 
\begin{figure}[ht]
\centering\includegraphics[scale=0.5]{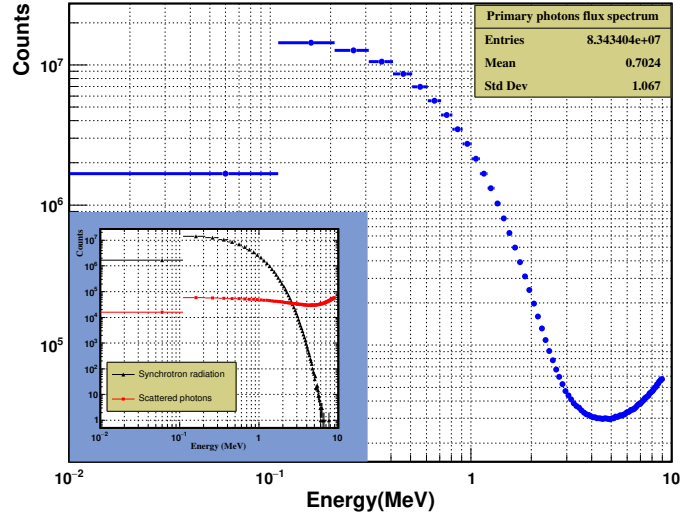}
\caption{The blue circle is the input sources for Geant4 simulation, including the scattered photons and the synchrotron radiation.  The inset shows these two parts respectively, the black triangle represents the synchrotron radiation from Figure 8(b), the red rectangle represents the energy spectrum of the scattered photons from Figure 9. \label{fig:The input photons flux} }
\end{figure}
There are three main interactions between gamma rays and matter, including photoelectric absorption, Compton scattering and electron pair production\cite{gamma}. The three interactions compete with each other and may exist at the same time. The relative importance of the three interactions is related to the gamma ray energy and the atomic number of the matter. At the same time, the attenuation coefficient of photons decreases exponentially with the material thickness. The Geant4 simulation is to change the thickness and type of materials to absorb photons, especially low-energy synchrotron radiation photons. Only polyethylene or if the layer is thinner, a large number of low-energy photons (tens of keV to hundreds of keV) will escape to cover the scattered photons. If the layer is thicker, the scattered photons will not pass through the shielding layer. Therefore, the high-density lead with appropriate thickness is added behind it for absorption. Through multiple simulations, it is concluded that the combination of $400\mathrm{cm}$ polyethylene and $0.2\mathrm{cm}$ lead have the best shielding effect. The transverse dimension of the shielding material is 120$\mathrm{cm}$. Table II shows the number of synchrotron radiation and scattered photons before and after shielding. B is the signal-to-noise ratio. After shielding, low-energy synchrotron radiation photons are absorbed. The synchrotron radiation photons generated by the bending magnet does not affect the detection of the scattered photons. Figure 11 shows the flux spectrum of photons after the lead target. 
\begin{table}[h]
\centering
\begin{tabular}{ccc}
\hline
Energy($0-9\mathrm{MeV}$) & Before shielding  & After shielding  \tabularnewline
\hline
scattered photons(S) & 3.519913$\times10^{6}$ & 648 
\tabularnewline

synchrotron radiation(N) & 7.991446$\times10^{7}$ & 2 
\tabularnewline

B(S/N)  & 0.44 & 324 \tabularnewline
\hline 
\end{tabular}
\caption{The number of synchrotron radiation and scattered photons before and after shielding. B is the signal-to-noise ratio. After shielding, low-energy synchrotron radiation photons are absorbed.}
\end{table}
\begin{figure}[ht]
\centering\includegraphics[scale=0.5]{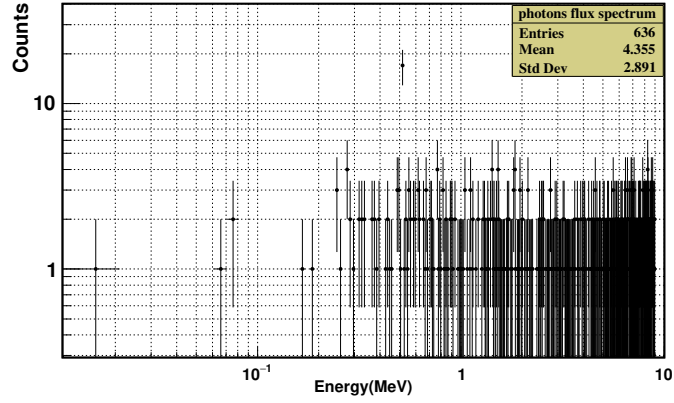}
\caption{ The photons flux spectrum is obtained after passing through the shielding materials. The number of photons is reduced to single digits and enters a HPGe detector. After a certain period of data accumulation, the maximum energy of the scattered photon is obtained by fitting the Compton edge. \label{fig:The flux of two gamma sources after passing through the shielding material} }
\end{figure}
The number of photons is reduced to single digits and enter a HPGe detector, which has an effective diameter of 5-6 $\mathrm{cm}$ and an effective height of 5-7 $\mathrm{cm}$\cite{2008The}. The detector is connected to a multi-channel analyzer. After a certain period of data accumulation, the maximum energy of the scattered photon is obtained by fitting the Compton edge.

\subsection{Possible background}
At the resonance frequency in a microwave resonant cavity, the maximum value of the energy storage of the electric field is equal to that of the energy storage of the magnetic field\cite{Microwave}. The energy storage in the cavity is 
\begin{equation}
W=\frac{\varepsilon_{0}}{2}\cdot 2\pi lE_{m}^{2}\int_{0}^{R} J_{0}^{2}(\frac{2.405}{R}r)rdr=\frac{1}{2}\pi\varepsilon_{0}R^{2}lE_{m}^{2}J_{1}^{2}(2.4),
\end{equation}
where the radius $R=11.65\mathrm{mm}$, the length of the cavity $l<2.1R=20\mathrm{mm}$, $E_{m}=1\times 10^{7}\mathrm{V/m}$. The energy storage can be calculated about $0.001\mathrm{J}$.
The quality factor of the cylindrical resonant cavity is
\begin{equation}
Q=\frac{R}{\delta(1+\frac{R}{l})},
\end{equation}
where the $\delta=\sqrt{\frac{2}{\omega\sigma\mu_{0}}}$ is the skin depth on the cavity wall. The conductivity $\sigma =5.8\times10^{7}\mathrm{S/m}$.
The electromagnetic field has losses on the metal wall of the resonant cavity. The quality factor Q can be expressed as $\omega W/P_{d}$. According to the Eq.(22) and Eq.(23), the power loss $p_{d}$ can be calculated about $6 \mathrm{kW}$. The external excitation source can be fed with the same power, so that the loss and supplementation can reach a dynamic balance. At this time, the stored energy inside the resonant cavity is stable. The stability of the field is conducive to the Compton backscattering process. 
With Eq.(23), the quality factor Q of the resonant cavity is obtained. Table III shows the resonance frequency and Q value of the cavity comes from the theoretical calculation, the CST simulations without holes and the CST simulations with the hole radius, $0.15\mathrm{mm}$. The results of the simulation is basic anastomotic with the theoretical calculations. The influence of the holes with the radius of $0.15\mathrm{mm}$ on the resonant frequency and the quality factor Q can be ignored. Table IV shows the variation of the resonance frequency in the cavity with the hole radius $1\mathrm{mm}$, $1.5\mathrm{mm}$, and $2\mathrm{mm}$. The resonance frequency changes at a smaller extent with the increase of the hole radius. In order to ensure that the resonance frequency is consistent with the theoretical value, the size of the resonant cavity can be slightly adjusted.
\begin{table}[h]
\centering
\begin{tabular}{ccc}
\hline
parameter & frequency(GHz)  & Q value  \tabularnewline
\hline
Theoretical calculation & 9.848975 & 11055.4 \tabularnewline

Simulation (without hole) & 9.848976 & 11048.2 \tabularnewline

Simulation (hole radius 0.15$\mathrm{mm}$) & 9.848973 & 11043.8 \tabularnewline
\hline 
\end{tabular}
\caption{The resonance frequency and Q value of the cavity comes from the theoretical calculation, the CST simulations without holes and the CST simulations with the hole radius, $0.15\mathrm{mm}$.}
\end{table}
\begin{table}[h]
\centering
\begin{tabular}{cc}
\hline
hole radius/mm & frequency/GHz   \tabularnewline
\hline
 1.0mm & 9.84790 \tabularnewline

 1.5mm & 9.84533 \tabularnewline

 2.0mm & 9.84026 \tabularnewline
\hline 
\end{tabular}
\caption{The variation of the resonance frequency in the cavity with the hole radius $1\mathrm{mm}$, $1.5\mathrm{mm}$, and $2\mathrm{mm}$. The resonance frequency decreases slightly with the increase of the hole radius. The influence on the resonance frequency can be corrected by fine-tuning the cavity size.}
\end{table} Figure 12 shows the normalized distribution of the field. The leaked field in the side wall is quickly cut off. With the increase of the hole radius, the distribution of the field has only a slight change in the local part of the hole. For the hole radius is 2$\mathrm{mm}$, there is a 10$\%$ difference between the local field of the hole and that of the standard resonant cavity. The influence on the field can be ignored for the small hole radius.

 \begin{figure}[ht]
\centering\includegraphics[scale=0.5]{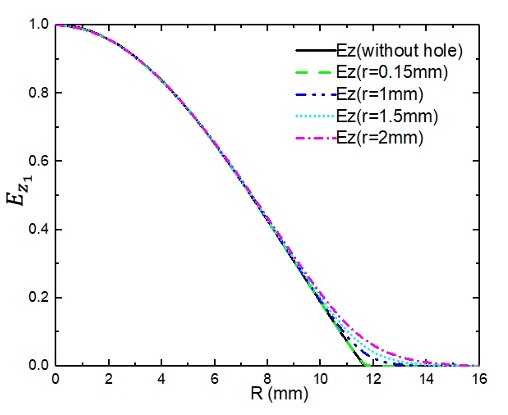}

\caption{The normalized distribution of the field in the direction of the cavity radius. The leaked field close to the wall is quickly cut off. With the increase of the hole radius, the field distribution has only a slight changes in the local space around the hole. The influence on the field can be ignored for the small hole radius.\label{fig:The normalized distribution of the field in the radius of the cavity} }
\end{figure}

In the strong electric field, the trajectory of the electrons changes. In the resonant cavity, the maximum field strength $E_{m}=1\times 10^{7}\mathrm{V/m}$, the average value $\bar E_{z_{1}}=6.11351\times10^{6}\mathrm{V/m}$ can be calculated by
\begin{equation}
E_{z_{1}}=E_{m}J_{0}(K_{c}r_{1}),
\end{equation}
\begin{equation}
\bar E_{z_{1}}=\frac{\int_{-R}^{R}E_{z_{1}}dr_{1}}{2R}.
\end{equation}
The influence can be estimated as follows:
\begin{equation}
	F=q\bar E_{z_{1}}=\frac{\gamma m_{0}c^{2}}{r}.
\end{equation}
The effective deflection radius $r=19.629\mathrm{km}$ for the electron energy of 120GeV. Therefore, with Eq.(21), the characteristic energy is $195.257\mathrm{keV}$. The influence of the magnetic field in the cavity can also be calculated, and the effective deflection radius $r=28.837\mathrm{km}$. The classical radiations induced by this electromagnetic field in the resonant cavity and the synchrotron radiations is of the same magnitude.  

\section{Error Analysis \label{sec:Error Analysis}}
The measurement uncertainty of the electron beam with the energy of 120$\mathrm{GeV}$ in the Higgs mode on CEPC is required to be smaller than $10 \mathrm{MeV}$. The resonant cavity is placed vertically in the beam tube. The Poynting vector is in the radius direction and also represents the motion direction of the microwave photons. The collision angle $\alpha$ is determined by the spatial alignment of the resonant cavity. The process of the spatial alignment is calibrated by a laser pulse. The laser pointing stability is up to $5\times 10^{-7}\mathrm{rad}$ \cite{laser}. The stability of the high-frequency microwave source can reach $10^{-5}\sim10^{-6}$. For $\omega_{max}^{\prime}=9\mathrm{MeV}$, $\omega_{0}=4.08032\times10^{-5}\mathrm{eV}$, the $\delta\omega_{0}\approx4.08\times10^{-10}\mathrm{eV}\sim4.08\times10^{-11}\mathrm{eV}$. Therefore, the main part that affects the accuracy of the beam-energy measurement is the energy resolution of the HPGe detector. If the HPGe detector is well calibrated, the detection uncertainty of the maximum energy of the scattered photons can reach $10^{-4}$. The uncertainty of the beam energy measurement $\delta\varepsilon_{0}$ is $6\mathrm{MeV}$ with Eq.(7). The relative error is $\delta\varepsilon_{0}/\varepsilon_{0}=5\times10^{-5}$.

\section{Summary\label{sec:Summary}}
In order to simplify the measurement system of the high energy beam energy, a new microwave-electron Compton backscattering method is proposed. The beam energy is obtained by detecting the energy distribution of the scattered photons and fitting the Compton edge. A cylindrical resonant cavity with $TM_{010}$ mode is selected to store the energy of the microwave photons. In a head-on collision mode, the microwave with the wavelength of $3.04\mathrm{cm}$ is chosen to collide with the electrons with the energy of $120\mathrm{GeV}$ in the Higgs mode on CEPC. The maximum energy of the scattered photons is $9\mathrm{MeV}$. The HPGe detector is the only choice and is employed to detect the energy distribution of the scattered photons. With our method, the measurement uncertainty of the beam energy is $6\mathrm{MeV}$. With the help of the Monte Carlo simulation, a shielding structure with $400\mathrm{cm}$ polyethylene and $0.2\mathrm{cm}$ lead is designed at the end of the beam line to minimize background noises. The simulation of CST software shows that the influence of the hole radius on the cavity field can be ignored. The cavity size can be fine-tuned to correct the change of the resonant frequency. The classical radiations from the electron beam in the resonant cavity is of the same magnitude of the synchrotron radiation.  

\section{acknowledgment}
This work is supported in part by National Natural Science Foundation
of China (11655003); Innovation Project of IHEP (542017IHEPZZBS11820,
542018IHEPZZBS12427); the CAS Center for Excellence in Particle Physics
(CCEPP); IHEP Innovation Grant (Y4545170Y2); Chinese Academy of Science
Focused Science Grant (QYZDY-SSW-SLH002); Chinese Academy of Science
Special Grant for Large Scientific Projects (113111KYSB20170005);
National 1000 Talents Program of China; the National Key Research
and Development Program of China (No.2018YFA0404300).

\section{data availability}
The data that support the findings of this study are available from the corresponding author upon reasonable request.

\end{spacing}

\begin{thebibliography}{}
\vspace{3mm}
\bibitem{Aad:2012tfa} Atlas Collaboration. "Observation of a new particle in the search for the Standard Model Higgs boson with the ATLAS detector at the LHC." arXiv preprint arXiv:1207.7214 (2012).
\bibitem{Chatrchyan:2012ufa} Chatrchyan, Serguei, et al. "Observation of a new boson at a mass of 125 GeV with the CMS experiment at the LHC." Physics Letters B 716.1 (2012): 30-61.
\bibitem{CEPCStudyGroup:2018ghi} CEPC Study Group. "CEPC Conceptual Design Report: Volume 2-Physics and Detector." arXiv preprint arXiv:1811.10545 (2018).
\bibitem{Arnaudon} Arnaudon, L., et al. "Accurate determination of the LEP beam energy by resonant depolarization." Zeitschrift für Physik C Particles and Fields 66.1 (1995): 45-62.
\bibitem{2020The} Tang, Guangyi, et al. "The circular electron–positron collider beam energy measurement with Compton scattering and beam tracking method." Review of Scientific Instruments 91.3 (2020): 033109.
\bibitem{ZHANG2019391} Zhang, J. Y., et al. "Energy deviation study of BEMS at BEPCII." Nuclear Physics B 939 (2019): 391-404.
\bibitem{Muchnoi} Muchnoi, Nickolai. "High precision beam energy measurement at VEPP-4M."
\bibitem{gamma-ray} An, Guang-Peng, et al. "High energy and high brightness laser Compton backscattering gamma-ray source at IHEP." Matter and Radiation at Extremes 3.4 (2018): 219-226.
\bibitem{1948Interaction} Feenberg, Eugene, and Henry Primakoff. "Interaction of cosmic-ray primaries with sunlight and starlight." Physical Review 73.5 (1948): 449.
\bibitem{2008The} Achasov, M. N. , et al. "The beam energy calibration system for the BEPC-II collider." Accelerator center (2008).
\bibitem{Microwave2} H.G. Tang, C.Q. Tian, "Microwave Technology principle and Application analysis.",China Science and Technology Investment 000.036(2012):156-156.
\bibitem{cross-section} Si M, Huang Y, et al. "The linear and nonlinear inverse Compton scattering between microwaves and electrons in a resonant cavity." arXiv preprint arXiv:2109.04213 (2021).
\bibitem{T1976} Suzuki, T. . "General Formulas of Luminosity for Various Types of Colliding Beam Machines." (1976).
\bibitem{SR} Xie Yaning, et al. "A calculation method of synchrotron radiation spectrum." Conference on Computational Physics. Chinese Nuclear Society, 2001.
\bibitem{sampling} Flury, Bernard D. "Acceptance–rejection sampling made easy." SIAM Review 32.3 (1990): 474-476.
\bibitem{gamma} Nelson, G., and Doug Reilly. "Gamma-ray interactions with matter." Passive nondestructive analysis of nuclear materials 2 (1991): 27-42.
\bibitem{Microwave} K.Y. Zhao, "Microwave principle and technology.", Higher Education Press(2006).
\bibitem{laser} F. Zhu, Y. Li, and J. Tan, Optical and Precision Engineering 28, 817 (2020).
\end{thebibliography}
\end{document}